\documentclass[fleqn,10pt]{wlscirep}
\usepackage[utf8]{inputenc}
\usepackage[T1]{fontenc}
\usepackage{lmodern}
\usepackage{csquotes}
\usepackage{amssymb}
\usepackage{algpseudocode}
\usepackage{enumitem}
\usepackage{amsmath}
\usepackage{mathtools}
\usepackage{physics}
\usepackage{xparse}
\usepackage{bm}
\usepackage{nicematrix}
\usepackage{amsthm}
\usepackage[inkscapepath=../assets/svg]{svg}
\usepackage{geometry}
\usepackage{caption}
\usepackage{subcaption}
\usepackage{titlesec}
\usepackage{hyperref}
\usepackage{rotating}
\usepackage{tikz}
\usepackage{longtable}
\usepackage{tabularx}
\hypersetup{
   colorlinks=true,
   linkcolor=blue,
   urlcolor=cyan
}
\theoremstyle{definition}

\usepackage{multirow}
\usepackage{geometry}

\usepackage{booktabs}
\usepackage{ragged2e}

\DeclareMathOperator*{\argmin}{arg\,min}

\title{QAL-BP: An Augmented Lagrangian Quantum Approach for Bin Packing}

\author[1*]{Lorenzo Cellini}
\author[2]{Antonio Macaluso}
\author[1]{Michele Lombardi}
\affil[1]{Department of Computer Science and Engineering, University of Bologna, Bologna, Italy}
\affil[2]{Agents and Simulated Reality Department, German Research Center for Artificial Intelligence (DFKI), Saarbruecken, Germany}

\affil[*]{corresponding author(s): Lorenzo Cellini (lorenzo.cellini3@studio.unibo.it)}

\begin{abstract}

The bin packing is a well-known NP-Hard problem in the domain of artificial intelligence, posing significant challenges in finding efficient solutions. Conversely, recent advancements in quantum technologies have shown promising potential for achieving substantial computational speedup, particularly in certain problem classes, such as combinatorial optimization.

In this study, we introduce QAL-BP, a novel Quadratic Unconstrained Binary Optimization (QUBO) formulation designed specifically for bin packing and suitable for quantum computation. QAL-BP utilizes the Augmented Lagrangian method to incorporate the bin packing constraints into the objective function while also facilitating an analytical estimation of heuristic, but empirically robust, penalty multipliers. This approach leads to a more versatile and generalizable model that eliminates the need for empirically calculating instance-dependent Lagrangian coefficients, a requirement commonly encountered in alternative QUBO formulations for similar problems.
To assess the effectiveness of our proposed approach, we conduct experiments on a set of bin packing instances using a real Quantum Annealing device. Additionally, we compare the results with those obtained from two different classical solvers, namely simulated annealing and Gurobi. The experimental findings not only confirm the correctness of the proposed formulation, but also demonstrate the potential of quantum computation in effectively solving the bin packing problem, particularly as more reliable quantum technology becomes available.

\end{abstract}
\begin{document}

\flushbottom
\maketitle

\section*{Introduction}

Bin packing is a well-established \cite{npstrongness} combinatorial optimization problem with wide-ranging applications in domains such as logistics, resources allocation, and scheduling. Its primary objective is to minimize the number of fixed-capacity bins required to pack a set of items of variable size.

Despite extensive research efforts, bin packing remains a challenging problem due to the exponential growth of solution possibilities as the number of items and bins increases \cite{DELORME20161}. On the other hand, quantum computing has recently emerged as a promising alternative to solving various AI problems, including coalition formation in multi-agent systems \cite{venkatesh2023gcs,venkatesh2022bilp, venkatesh2023quacs} and supervised learning \cite{macaluso2023maqa,macaluso2022variational}, although a real practical quantum advantage has yet to be found considering near-term quantum technology. 
The standard approach in quantum computing for optimization involves reformulating the original problem as a Quadratic Unconstrained Binary Optimization (QUBO) problem and employing quantum annealers (QAs) or parametrized quantum circuits, such as QAOA\cite{farhi2014quantum}, to find the optimal solution.
These approaches possess distinctive strengths and weaknesses. QAOA, for example, enables theoretical solutions to any QUBO problem with arbitrary precision by increasing the depth of the associated quantum circuit. On the other hand, QAs are specifically designed to identify the lower energy state of an Ising Hamiltonian representing the original QUBO problem and are better suited for tackling larger problems in terms of the number of QUBO variables.

In case of constraint optimization problems, the main drawback of the reformulation as a QUBO consists of associating a penalty term to the constraints and including them in the objective function. This approach requires empirical estimation of the penalty terms, which translates to running the QUBO solver (QA or QAOA) multiple times before achieving a feasible solution and poses several limitations to its real world applicability, especially when considering large problem instances and the imperfections of near-term quantum technology.

This paper presents a novel method for solving the bin packing problem using quantum computation. Specifically, we introduce an analytical heuristic approach for estimating penalty multipliers based on the Augmented Lagrangian framework \cite{Hestenes1969}, which allows to obtain a complete QUBO formulation without requiring empirical, instance-based parametrization. To demonstrate the effectiveness of our proposed method, we conduct experiments using a real quantum annealer and compare the results with two different state-of-the-art classical baselines. 

\section*{Problem Formulation}
The bin packing problem (BPP) is a classic optimization task that involves packing objects of different sizes into containers, or bins, with a limited capacity. 
The goal is to minimize the number of bins needed to pack all the objects.
A mathematical formulation of the bin packing problem can be expressed as follows: given a set of $n$ items of given integer size (or weight) $w_j \ (j = 1, \ldots , n)$ the goal is to pack them
 into the minimum number of identical bins of integer capacity $C$.

Let $m$ be any upper bound on the solution value and let introduce $y_i, x_{ij}$ two sets of binary variables such that:
$y_i (i = 1, \ldots, m)$ takes the value $1$ if and only if bin $i$ is used in the solution
and $x_{ij} (i = 1, \ldots, m; j = 1, \ldots, n)$ takes the value $1$ if and only if item $j$ is packed into bin $i$.
A commonly adopted Integer Linear Programming (ILP) problem formulation is the following\cite{martellototh1990}:

\begin{align}
    \label{eq:bpp-model}
    \argmin_{x, y} \ & \sum_{i=1}^m y_i \\
    \text{s.t. }
    & \sum_{i=1}^m x_{ij} = 1 & \forall j = 1, \dots, n \label{cnst:one_item} \\
    & \sum_{j=1}^n w_j x_{ij} \leq C y_i & \forall i = 1, \dots, m \label{cnst:capacity} \\
    & x_{ij} \in \{0, 1\} & \forall i = 1, \dots, m, \ \forall j = 1, \dots, n \nonumber \\ 
    & y_{i} \in \{0, 1\} & \forall i = 1, \dots, m \nonumber
\end{align}

A practical variant of significant interest is the \emph{online} bin packing problem. In this scenario, items of varying sizes are observed sequentially, and the decision maker must determine whether to select and pack the currently observed item or let it pass. Each decision is made without the ability to recall previous decisions.
In contrast, the \emph{offline} bin packing problem allows for rearranging the items in an attempt to achieve a better packing arrangement when additional items arrive. However, this approach necessitates additional storage to hold the items that need to be rearranged.

\section*{Related Works}

Classical algorithms for solving the BPP rely on Linear Programming relaxations and dynamic programming \cite{Eisemann1957TheTP, Gilmore}. However, as the number of items increases, the problem becomes intractable, and even for medium-sized instances, the optimal solution cannot be computed within a reasonable time frame. 
For this reason, several approximation algorithms and heuristics approaches can be adopted, such as simulated annealing\cite{Kampke, KokHua}, Tabu search\cite{Scholl1997}, population-based algorithms\cite{VAHRENKAMP1996191}, evolutionary and genetic heuristics\cite{Burke, Falkenauer, FalkenauerDelchambre, Quiroz, Sim} with hyper-heuristics\cite{Bai, LpezCamacho2011AHF, Sosa, Hart}, variable neighborhood search meta-heuristics\cite{Gomez} and ad-hoc crafted heuristics\cite{Eilon, Gupta, Lewis}.

In addition 
various alternative reformulations of the BPP have been proposed to improve the computational performance, such as pseudo-polynomial models \cite{Rao1976,Dyckhoff,Stadtler}. 
While these approaches offer more efficient problem formulations and enable the implementation of solutions that do not scale exponentially with the input size, they suffer from the drawback of the number of variables depending on both the number of items and the bin capacity. 

More recently, the adoption of quantum computing has been explored for solving BPP. Existing quantum solutions involve reformulating the original problem as a QUBO problem and leveraging gate-based quantum computers or quantum annealing.

At the time of writing this paper, two end-to-end QUBO models have been proposed for BPP, namely the Pseudo-Polynomial formulation \cite{Lodewijks} and the Unbalanced Penalization approach \cite{Barrera}. Alternatively, another existing approach \cite{hybrid_bpp} addresses the BPP through a hybrid approach, using quantum annealing to solve the sub-problem of filling a single bin, with the chance of reaching a sub-optimal solution.  

\paragraph{Pseudo-Polynomial QUBO formulation}\label{PP QUBO}

The Pseudo-Polynomial QUBO formulation for the BPP \cite{Lodewijks} is defined by three sets of binary variables. These variables are employed to represent the placement of weights in bins, indicate whether bins are empty or not, and specify the filling levels of the bins.
The corresponding Hamiltonian of the QUBO formulation consists of two weighted components, necessitating empirical estimation for the weights for each problem instance to avoid infeasible solutions. The primary objective of the Hamiltonian is to minimize the number of used bins, which constitutes the classical objective function.
The constraints Hamiltonian comprises three components. Firstly, it enforces the condition that each bin must be filled up to a unique level while ensuring that unused bins remain unfilled. Secondly, it guarantees that every item is allocated to a bin. Lastly, the third component penalizes configurations in which bins are overfilled, thereby violating the capacity constraint. Additionally, an extra term is introduced to account for considerations regarding only non-empty bins.

A significant drawback of this formulation is its limited scalability concerning the number of binary variables. Although the introduction of slack variables enables the formulation to be pseudo-polynomial, it results in the addition of $nC$ binary variables, making the formulation dependent on the specific problem instance due to the influence of bin capacity on the variable count.
This scalability issue becomes particularly pertinent in the context of modern Quantum Processing Units (QPUs), which face restrictions in handling relatively small problem instances due to qubit topology and connectivity constraints \cite{Kuramata_2021}. 
Consequently, even for small problem instances, a pseudo-polynomial Hamiltonian may become intractable when implemented on such quantum platforms.

Moreover, achieving a well-balanced weight assignment for each term in the Hamiltonian is of paramount importance to effectively minimize the number of used bins while satisfying the defined constraints. This requires to run the same problem instance multiple times using various hyperparameter sets in order to optimize the formulation's performance.

\paragraph{Unbalanced penalization formulation}
An alternative QUBO model for the BPP\cite{Barrera} introduces an inequality constraint $g(x) = \sum_i l_ix_i - C \le 0$ whose violation can be penalized using the exponential function $e^{g(x)}$. To ensure a valid QUBO model, the exponential function is expanded up to its second-order Taylor's term, resulting in the approximation $e^{g(x)} \approx 1+g(x)+\frac{1}{2}g(x)^2$.

Despite its improved efficiency compared to the pseudo-polynomial QUBO formulation, this work has several limitations. Firstly, this formulation requires the estimation of lambda parameters (that act as constraints multipliers) specific for each problem instance. This implies running the quantum algorithm multiple times to obtain feasible solutions for a single problem instance.

Secondly, the model's performance is evaluated on a limited set of problem instances, raising concerns about its generalizability to other instances. The model's scalability across instances with varying numbers of items is not demonstrated; the evaluation is restricted to randomly generated instances with the same number of items.

Furthermore, the experimental testing of the solution relies on QAOA, which poses restrictions on the number of problem variables due to the challenges associated with simulating even small-scale quantum systems.

\section*{Contribution}

In this work, we present QAL-BP (\emph{Quantum Augmented Lagrangian method for Bin Packing}), a novel QUBO formulation for the BPP based on the Augmented Lagrangian method. QAL-BP is an end-to-end method for deriving a QUBO formulation for the BPP that enables efficient scaling of logical qubits and the analytical estimation of the Lagrangian penalty terms. Specifically, we establish a connection between QUBO models and Augmented Lagrangian methods, leveraging advancements in both fields and fostering potential future synergies.

The proposed formulation offers several advantages. Firstly, it exhibits independence of the number of variables from the bin capacity. This eliminates the need for introducing slack variables, which typically increase the number of logical qubits and make quantum solutions infeasible for execution on real quantum hardware. 
Secondly, we analytically determine the Lagrangian penalty terms for a specific class of instances, avoiding running the quantum annealing multiple times on the same instance, as it is usually necessary for alternative approaches. Through experiments conducted on a real quantum annealing device, we demonstrate the effectiveness of our proposed approach.
Thirdly, we compare the performance of QAL-BP with state-of-the-art classical approaches. The results demonstrate that QAL-BP consistently yields feasible solutions, and in most cases, it leads to the global minimum.

To the best of our knowledge, this marks the first instance of an end-to-end analytical quantum solution for the BPP that has been rigorously tested across a diverse set of problem instances, displaying superior performance in comparison to existing quantum solutions. Furthermore, our results indicate promising potential concerning state-of-the-art classical solvers, particularly when more reliable quantum devices will become available.

\section*{Methods}\label{sec:QAL-BP}


\subsection*{QAL-BP: Quantum Augmented Lagrangian method for Bin Packing}

Augmented Lagrangian methods are a class of algorithms used to solve constrained optimization problems by incorporating the constraints into the objective function through penalty terms. Consider a constrained minimization problem of the form:
\begin{align}
\begin{split}
    \text{min} \ & f(\textbf{x}) \\
    \text{s.t. } & c_{i}(\textbf{x}) = \textbf{b} \ \forall i \in \mathcal{D}
\end{split}
\end{align} 
where $x$ is a candidate solution, $c_{i}(\textbf{x}) = \textbf{b}$ are a set of equality constraints and $\mathcal{D}$ is the set of constraints. 
The Augmented Lagrangian method consists in defining an unconstrained problem of the form:
\begin{equation}
    \label{al_method}
    \text{min} \ \Phi(\textbf{x}) = f(\textbf{x}) + \sum_{i \in \mathcal{D}}\rho_i(c_{i}(\textbf{x}) - \textbf{b})^2 + \sum_{i \in \mathcal{D}}\lambda_i(c_{i}(\textbf{x}) - \textbf{b})
\end{equation}

where $\rho_i$, $\lambda_i$ for \{i = 1, \dots , $|\mathcal{D}$|\} are the lagrangian multipliers.

In practice, when using the Augmented Lagrangian approach, it is common to introduce additional constraints that do not alter the set of feasible solutions but aid in faster convergence of the solver\cite{bertsekas1982constrained}.

For the BPP, we introduce the following additional constraints:
\begin{equation}
\sum_{j=1}^n y_ix_{ij} = \sum_{j=1}^n x_{ij} \quad \forall i = 1, \dots, m \label{pnlty:redundant}
\end{equation}

This set of constraints imply that if a bin $i$ is not used ($y_i=0$), it cannot contain any items ($\sum_{j=1}^n x_{ij} = 0$).

Therefore, the Quantum Augmented Lagrangian method for Bin Packing (QAL-BP) embeds the constraints using the Augmented Lagrangian approach as follows:
\begin{align}
    \text{argmin}_{x, y} \ & \delta\sum_{i=1}^m y_i \label{model:QBPP}\\
    & + \sum_{i=1}^{m} \lambda_i \left( \sum_{j=1}^n w_j x_{ij} - c_i y_i \right) \label{pnlty:linear_capacity}\\
    & + \sum_{i=1}^{m} \rho_i \left( \sum_{j=1}^n w_j x_{ij} - c_i y_i \right)^2 \label{pnlty:quadratic_capacity}\\
    & + \theta \sum_{j=1}^n \left(\sum_{i=1}^m x_{ij}-1\right)^2 \label{pnlty:one_item}\\
    & +  \gamma \sum_{i=1}^m \left(1 - y_i \right)\sum_{j=1}^n x_{ij} \label{pnlty:gamma_term}\\
    \text{s.t. }
    x_{ij}& \in \{0, 1\} \qquad \forall i = 1 \dots, m, \forall j = 1 \dots, n \nonumber \\ \nonumber
    y_{i} & \in \{0, 1\} \qquad \forall i = 1 \dots, m \\ 
    \delta, &\lambda, \rho, \theta, \gamma \ge 0 \nonumber
\end{align}

The penalties \eqref{pnlty:linear_capacity} and \eqref{pnlty:quadratic_capacity} represent the Augmented Lagrangian expansion of \eqref{cnst:capacity}. These terms impose a penalty of $\lambda_i s_i + \rho_i s_i^2$ for infeasible configurations while providing a negative reward to the solver for feasible configurations. Accurate estimation of $\lambda_i$ and $\rho_i$ values is crucial for correctly modeling the solution space.

Similarly, in Eq. \eqref{pnlty:one_item}, $\theta$ represents a penalty for not placing an item $j$, and penalizes $(k-1)\theta$ when item $j$ is placed $k$ times. Notably, this penalty term is not an Augmented Lagrangian expansion of \eqref{cnst:one_item}, but rather a pure squared penalty. This is because we do not wish to reward the solver when an item is not placed at all.

Finally, the term \eqref{pnlty:gamma_term} represents the penalty associated with the redundant constraints \eqref{pnlty:redundant} which imposes a penalty of $\sum_{j\in \mathcal{J}} x_{ij}$ when the set of items $\mathcal{J}$ is assigned to bin $i$ without setting the corresponding $y_i$ to 1.

It is important to note that the standard Augmented Lagrangian approach typically transforms inequality constraints into equality constraints using slack variables, which are then incorporated into the Lagrangian as shown in \eqref{al_method}. However, the proposed QUBO formulation in this study does not involve slack variables but directly utilizes the capacity constants $c_i$. This aspect provides a significant advantage over the pseudo-polynomial approach\cite{Lodewijks}.

\subsection*{Penalties estimation}\label{par:pen_estimation}

When incorporating constraints into the objective function, the estimation of penalty multipliers is typically carried out by testing a large set of parameters which requires running the algorithm multiple times with different parameter values in order to find an optimal solution for the specific problem instance, thus dramatically increasing the cost (in terms of time) to find good solutions.

Given the QAL-BP formulation and its corresponding set of constraints, we propose an analytical estimation of the penalty multipliers. The conditions are designed based on approximate worst-case reasoning, aiming to yield optimal or slightly sub-optimal solutions for most instances of the BPP.
The following is a set of heuristic conditions that consider each penalty multiplier individually, with the exception of the pair $\lambda, \rho$.
Considering the $i$-th bin, the correspondent Augmented Lagrangian term is given by $\lambda_i (s_i - c_i y_i) + \rho_i (s_i - c_i y_i)^2$. 
When $y_i = 0$, i.e, when the bin $i$ is not included in the final solution, using the smallest bin usage amount should be at least as expensive as using the bin:
\begin{equation}
    \lambda_i (w_{min} - 0) + \rho_i (w_{min} - 0)^2 \geq 1, \label{eq:first_cond}
\end{equation}
where $w_{min}$ is the smallest item weight, i.e. $w_{min} = \min \{w_{j}\}$.
If this condition is satisfied, using more capacity makes the solver choose to set $y_i = 1$. On the contrary, if $y_i = 0$ and no capacity is used, then the condition in eq. \eqref{eq:first_cond}
is trivially true.
Let's consider now the case where $y_i = 1$, meaning that the bin $y_i$ is included in the candidate solution. In this case, exceeding the capacity by any amount should be at least as expensive as using one more bin, i.e.:
\begin{equation}
    \lambda_i (c_i+w_{min} - c_i) + \rho_i (c_i+w_{min} - c_i)^2 \geq 1
\end{equation}
which conveniently is the same condition as Eq. \eqref{eq:first_cond}. 
Lastly, it is necessary to identify a solution space that contains only feasible solutions. 
In this case, the Lagrangian term needs to provide a positive reward (i.e., negative cost) if the constraint is satisfied. It is also necessary for such a reward to be small enough that so it does not provide an incentive for using another bin, i.e.:
\begin{equation}
\label{eq:lambda_calib}
    \lambda_i (- \frac{c_i}{2}) + \rho_i (- \frac{c_i}{2})^2 \geq 0
\end{equation}
Basically, here we are fitting a quadratic function using the conditions \eqref{eq:first_cond} and \eqref{eq:lambda_calib} to approximate the solution space excluding infeasible solutions.
We can therefore obtain values for $\lambda_i$ and $\rho_i$ by stating all conditions for their least restrictive values:
\begin{align}
    w_{min}\lambda_i + w_{min}^2\rho_i & = 1; \label{eq:first_cal_cond}\\
    -\frac{c_i}{2} \lambda_i + \frac{c_i}{4}^2 \rho_i & = 0,
\end{align}
which leads to an analytical formulation of the form:
\begin{align}
    \lambda_i & = \frac{c_i}{w_{min}\left( 2w_{min} + c_i\right)} \\
    \rho_i & = \lambda_i\frac{2}{c_i}= \frac{2}{w_{min}\left( 2w_{min} + c_i\right)}
\end{align}
The next step is to calibrate $\theta_j$. The abstract Lagrangian term associated with item $j$ is $\theta_j(p_j-1)^2$, where $p_j \in \mathbb{N}$ is the number of times item $j$ has been assigned to a bin. At this stage, only the assignment is considered and not capacities, so it is possible to ignore the index $j$ when defining the $\theta$ parameter. 
Moreover, we want to force the solver to assign all items to at maximum one bin, so the penalty should increase when $p_j \ne 1$. Thus, in case $p_j=1$, the following condition holds:
\begin{align}
    \theta(1-1)^2 \ge -1.
\end{align}
In case $p_j=0$, i.e., item $j$ is not assigned to any bin, then the associated penalty needs to be greater than the cost of opening a new bin, i.e.,
\begin{align}
    \theta \ge 2.
\end{align}

Also, the parameter $\gamma$ needs to be set. The abstract Lagrangian term associated with the $\gamma$ term is
    $\gamma(1-y_i)k_i$, which comes into play only when $y_i=0$ and $k_i\ne0$ by adding $k_i$ times the penalty $\gamma$. We want the minimum penalty to be at least equal to the cost of opening a new bin, i.e., 
\begin{align}
    \gamma \ge 1.
\end{align}
The final parameter to be estimated is $\delta$. Although it is not strictly a penalty term since it serves as a multiplier of the objective function, including it is beneficial for controlling the other parameters. One purpose of $\delta$ is to prevent issues that may arise when working with very small numbers.

Another reason for incorporating this multiplier is to address the undesirable behavior of the model in certain item configurations. In some cases, the model may favor configurations where one or more bins are slightly overfilled due to the high cost associated with opening a new bin. Naturally, this behavior is contingent on the specific combination of instance weights relative to the bin capacity, as well as their number.

To rectify this behavior, the following requirements must be met:
\begin{align}
    \delta \le \lambda s_{min} + \rho s_{min}^2,
\end{align}
where $s_{min}$ is the minimum capacity that can be exceeded. Thus,  the cost of opening a new bin must be less than the cost of overfilling an already open bin, of the smallest possible amount, i.e., $s_{min} \ge 1$.


\subsection*{Model analysis}

When solving QUBO problems with quantum computing, the number of binary variables of the problem formulation corresponds to the number of logical qubits to use in the quantum computer. Therefore, having an efficient formulation that minimizes the number of variables without limiting the range of possible solutions is crucial for the adoption of quantum approaches. In terms of variable count, the QAL-BP model is more efficient than the pseudo-polynomial one \cite{Lodewijks}, and equivalent, with respect to the \emph{Unbalanced penalization formulation}\cite{Barrera}. Specifically, for a given problem instance $BPP(n, C)$, the total number of variables is equal to the number of bins $m$ plus $n \times m$ decision variables representing the assignment of a specific item to a specific bin. Thus, in the worst case, $m=n$,  resulting in $n(n+1)$ binary variables. 

Furthermore, from a methodological point of view, the QAL-BP approach offers a twofold advantage. Firstly, the number of binary variables is not affected by bin capacities and item weights, as observed in the pseudo-polynomial formulation\cite{Lucas_2014}. Secondly, the reduced number of variables enables the execution of the QUBO problem using a smaller set of logical qubits, making it suitable for current QPUs.

\begin{figure}[ht]
    \center
    \includegraphics[width=0.4\linewidth]{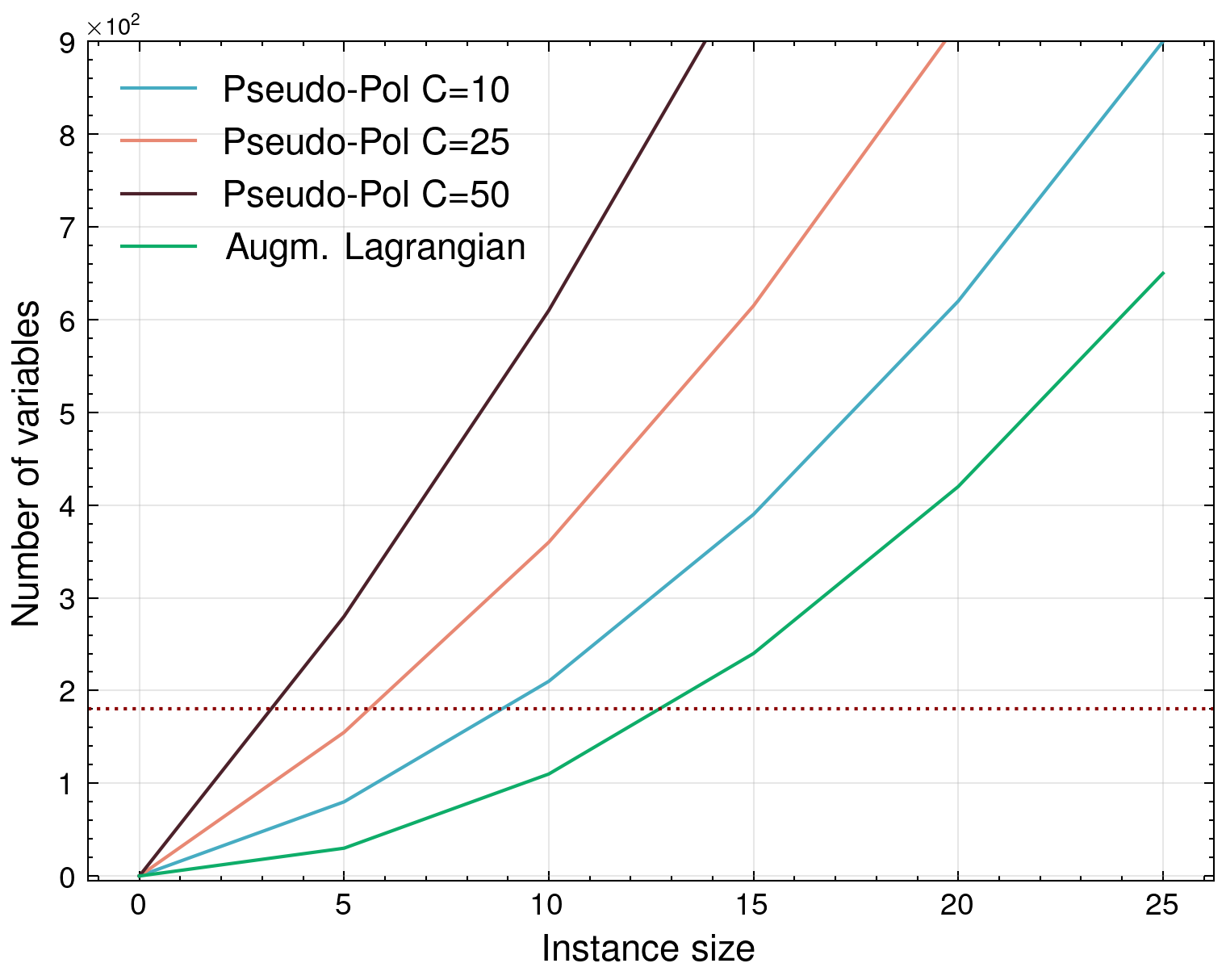}
    \caption{Comparative analysis of variable growth in the Pseudo-Polynomial and Augmented Lagrangian models concerning the number of items and bin capacity. Three distinct values of bin capacity (C) are explored. The continuous dark red line represents the upper limit for QUBO problems represented by fully connected graphs that can be mapped in the D-Wave Advantage Quantum Processing Unit (QPU) equipped with 5640 qubits.}
    \label{fig:num_vars_models}
\end{figure}

Fig. \ref{fig:num_vars_models} shows a comparison of the performance, in terms of the number of variables, between the QAL-BP and the pseudo-polynomial formulation. 

\section*{Evaluation}


\subsubsection*{Experimental Settings} \label{sec:data}

\paragraph{Data}
The experiments are performed on a set of eight classes of randomly generated instances, ranging from $3$ to $10$ items, with corresponding weights ranging from $4$ to $10$ and fixed bin capacity equal to $10$ (Table \ref{tab:dataset}). This choice allows exploring problem instances of different sizes while taking into account the limitations of modern QPUs which are restricted to non-sparse QUBO problems with up to $180$  binary variables\cite{Kuramata_2021}. 
Five different problem instances are generated for a fixed number of items, that vary in the weights. 
\begin{table}[ht]
    \centering
    \resizebox{.647\textwidth}{!}{%
    {\def\arraystretch{1.15}\tabcolsep=10pt
\begin{tabular}{ |c|>{\centering\arraybackslash}p{2cm}|>{\centering\arraybackslash}p{6cm}|>{\centering\arraybackslash}p{2cm}|>{\centering\arraybackslash}p{3cm}| } 
\hline
Seed & N. items & Weights & Lower bound& Instance name \\
\hline\hline
\multirow{8}{5em}{\centering  23} & 3 &  [4, 8, 6]                        &2 & (3, 23)\\ 
                      & 4 &  [8, 5, 4, 8]                     &3 & (4, 23)\\
                      & 5 &  [4, 4, 8, 8, 9]                  &3 & (5, 23)\\
                      & 6 &  [7, 5, 5, 5, 4, 9]               &4 & (6, 23)\\
                      & 7 &  [9, 7, 8, 6, 9, 6, 7]            &5 & (7, 23)\\
                      & 8 &  [4, 5, 7, 5, 6, 4, 6, 4]         &4 & (8, 23)\\
                      & 9 &  [7, 6, 8, 4, 8, 4, 9, 6, 4]      &6 & (9, 23)\\
                      & 10&  [5, 8, 6, 7, 10, 9, 4, 10, 7, 4] &7 & (10, 23)\\ 
\hline
\multirow{8}{5em}{\centering 42} & 3 &  [4, 8, 6]                        &2 & (3, 42)\\
                      & 4 &  [7, 7, 10, 4]                    &3 & (4, 42)\\
                      & 5 &  [8, 5, 4, 7, 10]                 &4 & (5, 42)\\
                      & 6 &  [9, 9, 9, 9, 7, 4]               &5 & (6, 42)\\
                      & 7 &  [9, 7, 7, 6, 5, 10, 9]           &5 & (7,42)\\
                      & 8 &  [8, 6, 9, 7, 7, 7, 5, 4]         &5 & (8, 42)\\
                      & 9 &  [7, 10, 4, 10, 9, 5, 8, 5, 9]    &7 & (9, 42)\\
                      & 10&  [8, 6, 4, 10, 7, 10, 8, 9, 9, 5] &7 & (10, 42)\\  
\hline
\multirow{8}{5em}{\centering 123} & 3 &  [4, 8, 8]                     &2 & (3, 123)\\ 
                      & 4 &  [4, 10, 5, 5]                  &3 & (4, 123)\\
                      & 5 &  [5, 6, 5, 6, 9]                &3 & (5, 123)\\
                      & 6 &  [7, 10, 7, 5, 9, 9]            &5 & (6, 123)\\
                      & 7 &  [10, 10, 4, 7, 5, 5, 5]        &5 & (7, 123)\\
                      & 8 &  [9, 9, 5, 6, 9, 5, 8, 7]       &6 & (8, 123)\\
                      & 9 &  [10, 9, 5, 9, 9, 5, 7, 9, 5]   &7 & (9, 123)\\
                      & 10&  [5, 5, 4, 7, 4, 8, 6, 5, 6, 4] &5 & (10, 123)\\  
\hline
\multirow{8}{5em}{\centering 90} & 3 &  [8, 6, 4]                        &2 & (3, 90)\\ 
                      & 4 &  [8, 5, 7, 6]                     &3 & (4, 90)\\
                      & 5 &  [6, 7, 8, 7, 4]                  &3 & (5, 90)\\
                      & 6 &  [7, 8, 9, 9, 10, 6]              &5 & (6, 90)\\
                      & 7 &  [6, 4, 4, 4, 8, 9, 6]            &4 & (7, 90)\\
                      & 8 &  [7, 10, 8, 8, 8, 5, 5, 8]        &6 & (8, 90)\\
                      & 9 &  [9, 6, 4, 10, 10, 5, 4, 4, 6]    &6 & (9, 90)\\
                      & 10&  [9, 6, 8, 7, 8, 10, 9, 6, 9, 10] &8 & (10, 90)\\ 
\hline
\multirow{8}{5em}{\centering 510} & 3 & [5, 8, 6]                       &2 & (3, 510)\\ 
                      & 4 &  [7, 9, 5, 5]                    &3 & (4, 510)\\
                      & 5 &  [6, 10, 4, 9, 4]                &3 & (5, 510)\\
                      & 6 &  [5, 5, 9, 10, 8, 6]             &4 & (6, 510)\\
                      & 7 &  [9, 7, 9, 4, 10, 10, 8]         &6 & (7, 510)\\
                      & 8 &  [9, 10, 8, 9, 4, 4, 9, 5]       &6 & (8, 510)\\
                      & 9 &  [5, 9, 10, 9, 7, 8, 4, 10, 6]   &7 & (9, 510)\\
                      & 10&  [10, 5, 9, 5, 8, 9, 7, 4, 6, 9] &7 & (10, 510)\\  
\hline
\end{tabular}
}
}
    \caption{
    Bin Packing problem instances. Each instance is characterized by specific parameters, which are organized into columns. The first column denotes the numeric seed utilized for generating the instances. The second column indicates the number of items that need to be placed within the bins. The third column comprises an array representing the weight of each individual item. The fourth column provides the $L_1$ lower bound for the given instance\cite{MARTELLO199059}. The last column assigns a unique label to facilitate identification and reference to each specific instance.}
    \label{tab:dataset}
  \end{table}

\paragraph*{Solving methods}



The quantum solver employed in this study is the \emph{D-Wave Advantage 4.1}, featuring a total of $5640$ physical qubits. 
For comparison, two classical solvers are utilized. The first is \textit{simulated annealing} (SA), which is considered the classical counterpart of quantum annealing, enabling a thorough assessment of the correctness of the QAL-BP formulation without encountering any errors inherent to real quantum hardware. Both quantum annealing and simulated annealing are available in the \textit{D-Wave Python library}\cite{DWave}.  In addition, we also solve the ILP formulation from Eq.~\eqref{eq:bpp-model}-\eqref{cnst:capacity} via the \textit{Gurobi} optimizer \cite{GurobiWebsite}, as a representative classical state-of-the-art method that relies on the branch-and-bound technique to efficiently find the optimal solution.

As previously mentioned, the rationale behind choosing quantum annealing over alternative quantum approaches, such as QAOA, is to enable a direct performance comparison between the most powerful and dependable current quantum technology and the state-of-the-art classical optimizer. This comparison aims to demonstrate the capability of current quantum computation in solving the BPP in relation to the best available classical solution. It is important to recognize that quantum hardware is still in its nascent stage, making such a comparative analysis critical in assessing the advancements and potential of quantum computing in tackling optimization problems like BPP.

\paragraph*{Metrics} 
In order to assess the performance of QAL-BP, we consider four distinct aspects: \emph{scalability}, \emph{runtime}, \emph{solution quality}, and \emph{feasibility ratio}. 

Regarding \emph{scalability}, our investigation encompasses a thorough examination of the necessary physical qubits for QAL-BP. 
In fact, when using a real quantum annealing device, a single logical variable (or logical qubit) in the QUBO problem is typically represented by a set of physical qubits within the real hardware, by means of chains. This is due to the incomplete connectivity of the QPU topology. The representation of a variable through a chain imposes the condition that all constituent qubits must take identical values for a given sample. Achieving this uniformity is contingent upon establishing robust couplings between the qubits along the connecting edges. Specifically, a strong coupling is implemented to ensure that the qubits in a chain are forced to return identical values and is controlled via a parameter called \emph{chain\_strength}. This metric is used to strengthen the coupling between qubits belonging to the same chain, against other topologically close qubits that may exert influence leading to divergent outcomes.
In this context, we examine the \texttt{chain\_break\_fraction}, which serves to quantify the fraction of chains in a sample that experiences disruption. A small \texttt{chain\_break\_fraction} is indicative of a more favorable outcome. Additionally, we implement different strategies to mitigate the chain break effect. Each strategy corresponds to different values of \emph{chain\_strength}. The first strategy, \emph{uniform\_torque\_compensation}, aims to compensate for the random torque of neighboring qubits that could potentially break the chain. The prefactor is set to the default value of $1.414$, resulting in a chain strength in the range $2.3$-$5.5$. The second strategy, \emph{scaled}, involves adjusting the chain strength to align with the problem bias, which is defined by the values taken by the linear and quadratic terms in the corresponding Ising problem. This calibration ensures a chain strength between $2.16$ and $16.5$. The third strategy, \emph{constant\_10}, selects a constant value of $10$ based on reasoning provided in a D-Wave whitepaper\cite{dwaveWhitePaper}. The fourth strategy, \emph{constant\_6}, employs a constant value of $6$, chosen by rounding up the maximum chain strength found by the \emph{uniform\_torque\_compensation} technique.

The \textit{runtime} is described in terms of Time-To-Solution (TTS), denoting the duration required by the solver to generate the ultimate solution, with measurements expressed in microseconds ($\mu$s). For QA, we sum up two different metrics: the \emph{qpu\_sampling\_time} and the time for \textit{embedding}.
The variable \emph{qpu\_sampling\_time} serves as an indicator of the active problem-solving duration of the QPU. 
Also, we incorporate the \textit{embedding} time, signifying the process of mapping an arbitrarily posed binary quadratic problem directly onto the QPU topology of an actual D-Wave system. Both \emph{qpu\_sampling\_time} and \textit{embedding} stand as the two primary components requiring consideration in evaluating QA, as their impact is contingent upon the specific hardware in use and the nature of the QUBO problem to be solved. For comprehensive coverage, we also present an in-depth description of all the metrics supplied by D-Wave, emphasizing that their influence is negligible when evaluating QA computations. 
Differently from QA, SA and Gurobi run locally. Thus, the TTS of SA is the time to run the annealing function, while for Gurobi, TTS is calculated as the time required to obtain the solution for a given problem instance.

The \emph{quality} of a solution is defined by the number of bins utilized in a given configuration, with a lower count of bins indicating a superior solution.

The \emph{feasibility ratio} serves as a quantitative measure of the probability that the solution with the minimum energy aligns with a feasible outcome. More precisely, this ratio is computed as the quotient between the number of instances in which the solution with minimum energy corresponds to a feasible solution (not necessarily the global optimum) and the total number of instances within each class of instances.
This metric holds significance exclusively within the context of simulated and quantum annealing, enabling an evaluation of the efficacy of contemporary quantum technology in providing solutions for the QAL-BP formulation.
  
\paragraph*{Models parameters} According to the analytical penalty estimation previously described, the multipliers are determined as follows: $\delta = 0.15; \ \lambda = 0.1389; \   \rho = 0.0278; \ \theta = 2; \  \gamma = 1$.

\subsection*{Results}

\subsubsection*{Scalability}\label{ssec:Scalability}

\begin{figure}[ht]
    \center
    \includegraphics[width=1\linewidth]{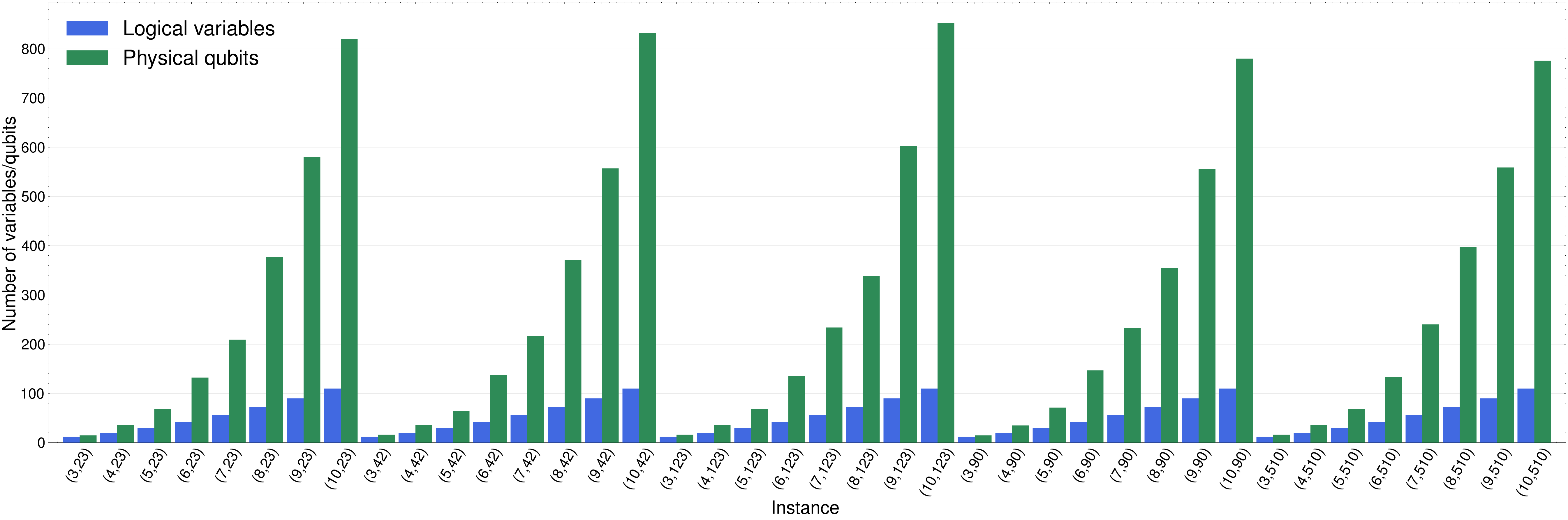}
    \caption{Comparison of the number of logical variables of QAL-BP against the physical qubits needed to implement the QUBO problem on the \emph{D-Wave Advantage 4.1}.}
    \label{fig:qubits}
\end{figure}

Figure \ref{fig:qubits} illustrates the required number of physical qubits for embedding BPP instances on a QPU. It is evident that the QAL-BP formulation necessitates an exponential increase in the number of physical qubits in comparison to the number of logical qubits (i.e., the number of variables in the QUBO). This characteristic represents a notable limitation when attempting to apply the formulation to real-world problems using existing quantum hardware.


\begin{figure}[ht]
    \center
    \includegraphics[width=.9\linewidth]{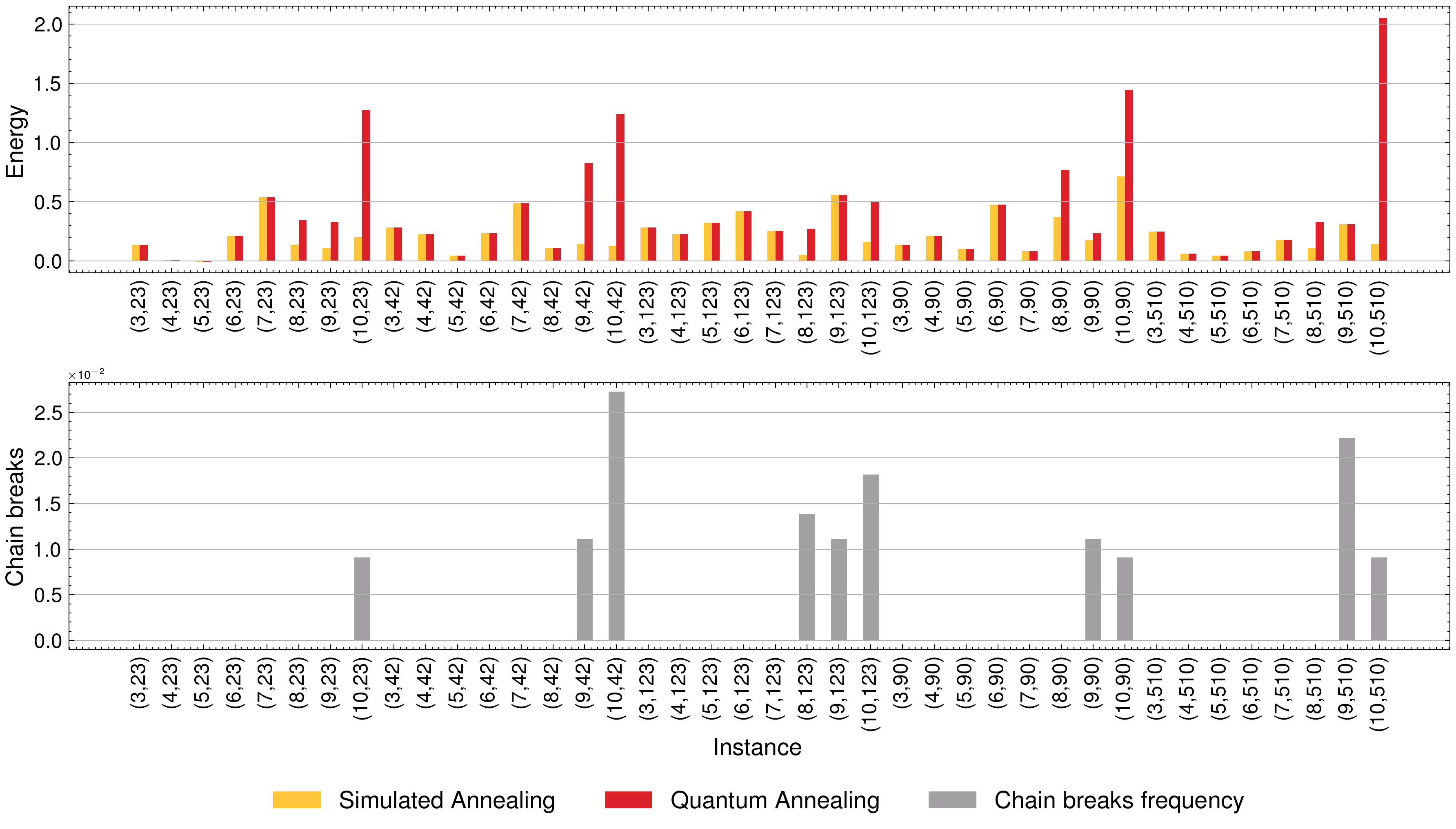}
    \caption{Comparative evaluation of the energy of the solutions provided by SA and QA (top), and the \texttt{chain\_break\_fraction} reported by the QA implementation (bottom). 
    }
    \label{fig:minimum_eigenvals}
\end{figure}

Additionally, the inherent limited connectivity of the QPU's topology introduces a vulnerability through coupling chains that connect the physical qubits encoding a single logical variable. Figure \ref{fig:minimum_eigenvals} offers insight into the distribution of the \texttt{chain\_break\_fraction} for each instance and its consequential impact on the solutions derived from QA.

For problem instances involving $3$ to $7$ items, the \texttt{chain\_break\_fraction} is null, and the QA yields identical solutions to SA. This implies that QA successfully identifies the global optimum within the QAL-BP formulation. Nevertheless, when the number of items equals or exceeds $8$, the energy associated with SA solutions consistently surpasses that of QA. This persists even when the \texttt{chain\_break\_fraction} is null for four out of five instances with $8$ items, although the observed difference is relatively small in this specific case.
Upon closer examination of chain breaks in instances involving more than $9$ items, the QA solutions markedly deteriorate compared to those produced by the SA. Notably, in all these cases, the \texttt{chain\_break\_fraction} consistently exceeds $0$. Hence, while the occurrence of chain breaks undeniably impacts solution quality, it is not the sole determinant preventing QA from returning to the global optimum, as evidenced by instances involving $8$ items.

\begin{figure}[ht!]
    \center
    \includegraphics[width=.85\linewidth]{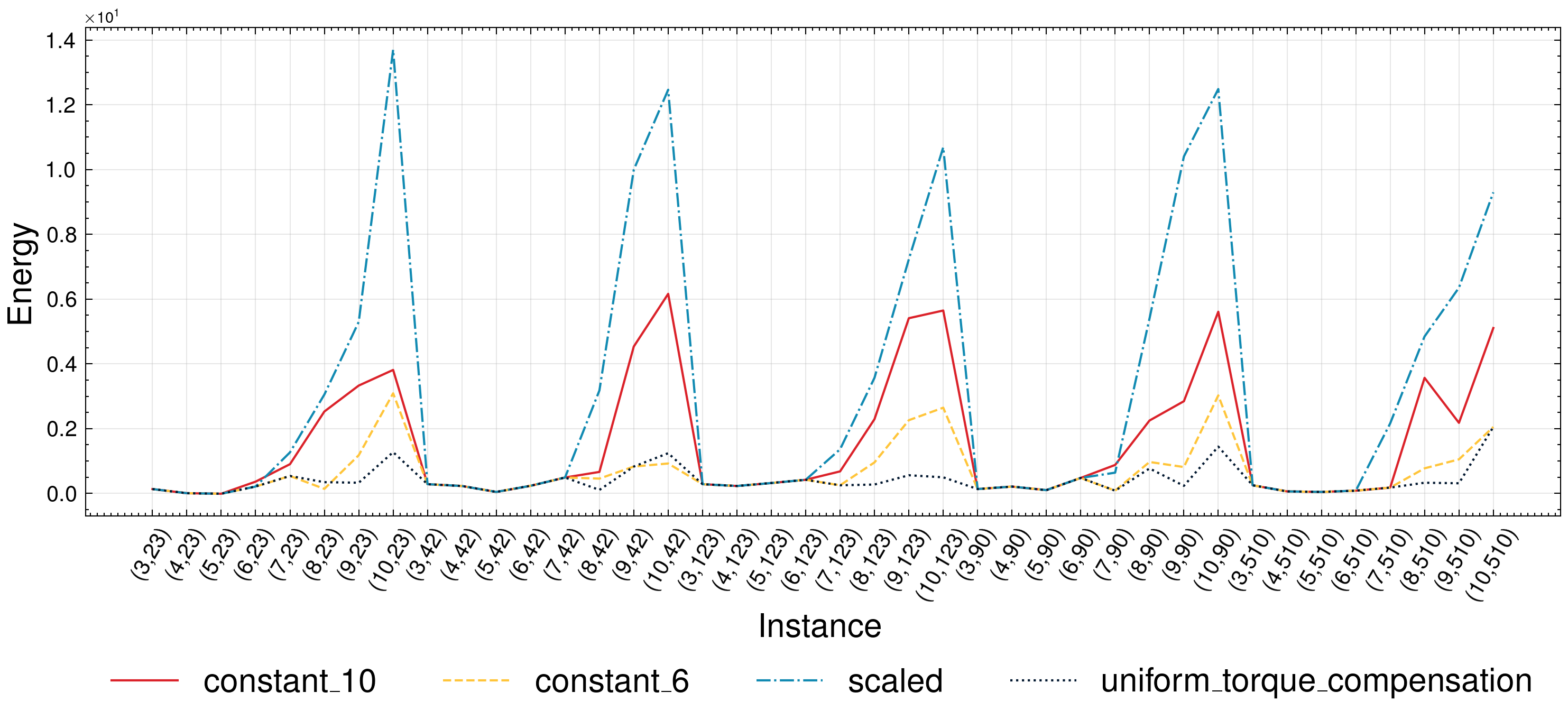}
    \caption{Comparative evaluation of the energy values for different chain break mitigation strategies.}
    \label{fig:energy_vs_chain_strength}
\end{figure}

The effect of chain breaks can be alleviated through the adjustment of \emph{chain\_strength} during the embedding of the QUBO problem into the topology of a real quantum annealer. To address this, we conduct a set of experiments testing four distinct strategies to determine the optimal \emph{chain\_strength} \cite{Carugno2022}. 
Results are shown in Figure \ref{fig:energy_vs_chain_strength}. 

We observe a correlation between the increase in chain strength and a corresponding increase in energy, potentially leading to a reduction in the number of successfully solved problems. 
This phenomenon is explained by the fact that when submitting a problem to the QPU, the auto-scaling feature scales all QUBO
weights in the range $[-1, +1]$. If the chain strength is too large, it causes 
individual QUBO coefficients, designed to control both the terms requiring minimization and the problem constraints, to shrink toward near-zero values. Consequently, each chain assumes an independent status, transforming the QUBO into a problem characterized by independent variables that lack mutual interactions\cite{dwaveWhitePaper}.

To identify the most effective mitigation strategies, we solved all instances previously described and calculated, for each, the percentage of cases where the optimal solution with minimum energy aligns with a feasible solution. The results are as follows: $52.5\%$ for \emph{scaled}, $57.5\%$ for \emph{constant\_10}, $67.5\%$ for \emph{constant\_6}, and $82.5\%$ for \emph{uniform\_torque\_compensation}.
Based on these results, we designate \emph{uniform\_torque\_compensation} as the most effective mitigation strategy for QAL-BP. Consequently, this strategy is employed in all subsequent experiments discussed throughout the remainder of this paper.

\subsubsection*{Runtime}

\begin{figure}[ht!]
\center
    \includegraphics[width=.7\linewidth]{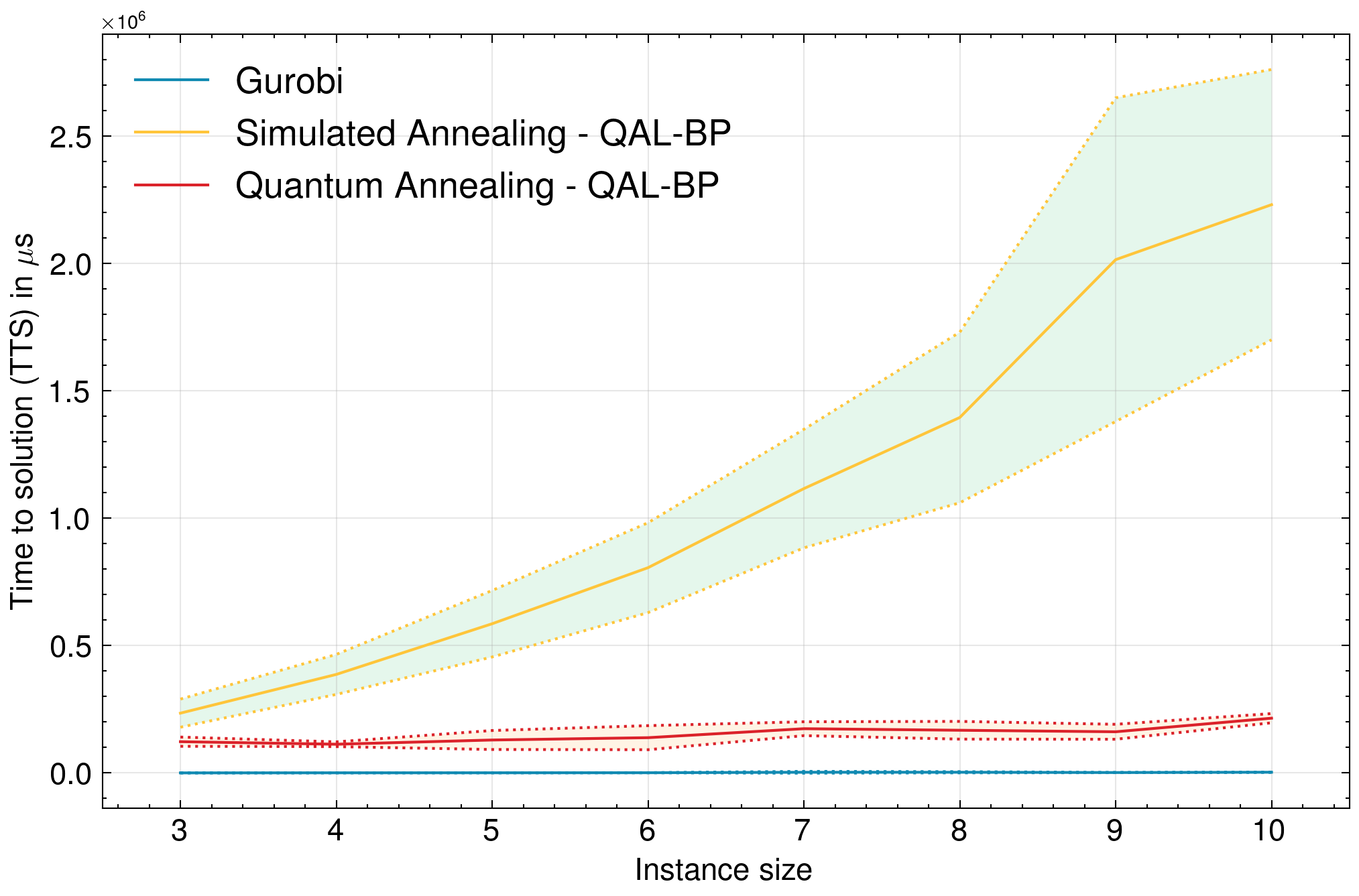}
    \caption{TTS comparison between Quantum Annealing (QA), Simulated Annealing (SA) and Gurobi. The Gurobi curve ranges between 440$\mu s$ and 3000$\mu s$.}
    \label{fig:tts}
\end{figure}

Figure \ref{fig:tts} illustrates the experimental results concerning TTS. The performance of SA deteriorates rapidly with the number of items, becoming extremely inefficient even for relatively small problems. 

In the comparative analysis between QA and Gurobi, the latter consistently exhibits superior performance, although the asymptotic behavior seems to be equivalent between the two methods. However, this might be due to the relatively small problem instances that the classical optimizer can easily solve. In fact, the Gurobi relies on an exact branch-and-bound approach \cite{MORRISON201679}, which in the worst case ends up exploring the entire solution space, resulting in prohibitive runtime when dealing with large problems.
Nevertheless, the QA has demonstrated the capacity to scale more efficiently than Gurobi for very large problems \cite{venkatesh2023q}, with a runtime linear in the input size \cite{venkatesh2022bilp, venkatesh2023gcs}. 
Taking these considerations into account, the findings suggest the potential for QA to surpass Gurobi's performance as more reliable quantum technology becomes available. Nonetheless, due to constraints in running larger problem instances, a definitive assessment of performance remains elusive.

\begin{figure}[ht]
    \centering
    \includegraphics[width=0.75\linewidth]{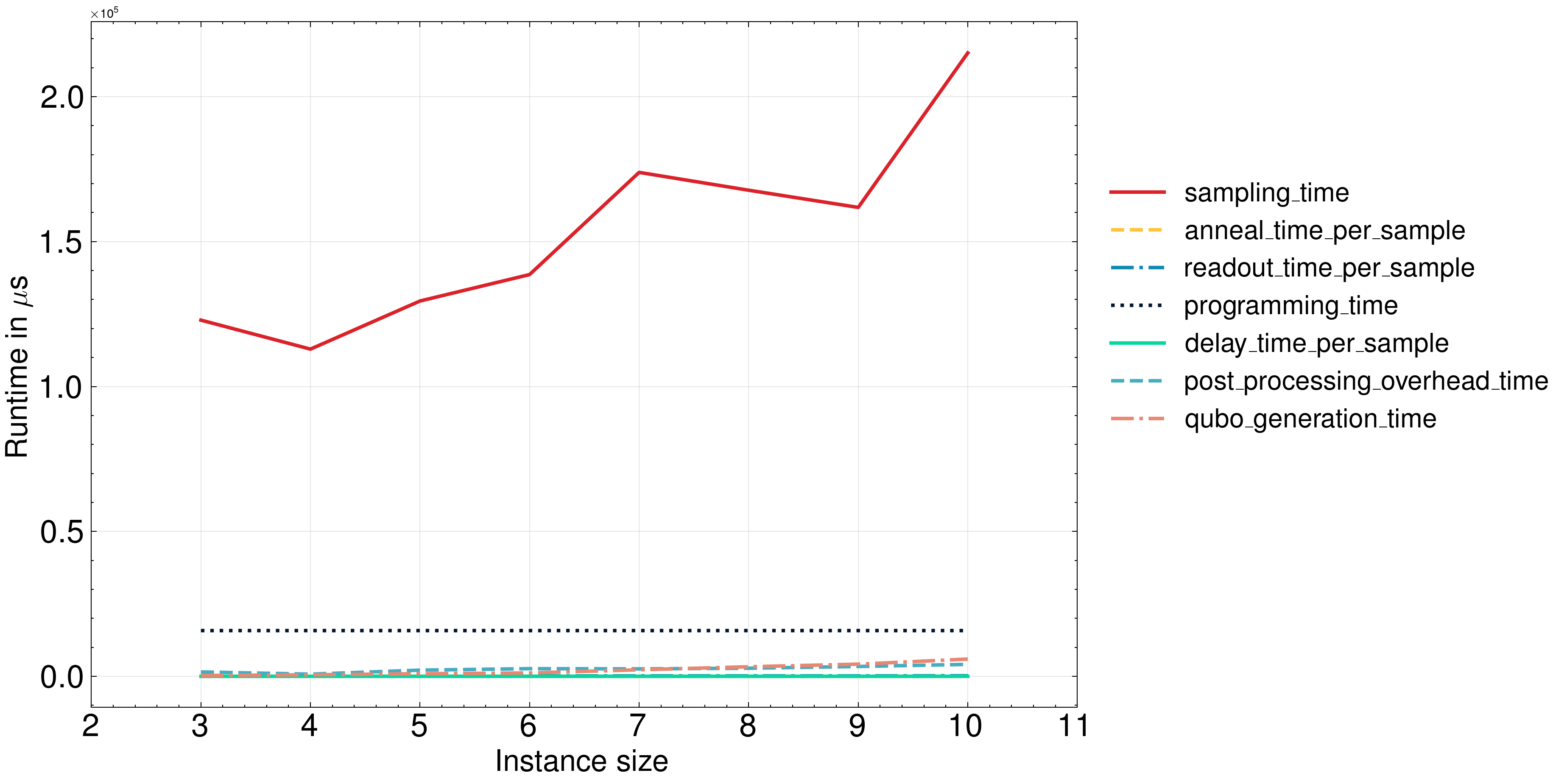}
    \caption{Runtime breakdown of the D-Wave QPU when solving the QUBO problem according to the QAL-BP formulation. Each metric is the average over all the instances of the same class, i.e., problems with the same number of input bins. In addition to the metrics provided by the D-Wave Python Library, we include the average time per instance class necessary to transform the ILP formulation into QAL-BP (\emph{qubo\_generation\_time}).}
    \label{fig:runtime_metrics}
\end{figure}

Furthermore, we present the averages across all problem instances for various metrics offered by the D-Wave platform\cite{DWave}. As depicted in Figure \ref{fig:runtime_metrics}, the TTS associated with annealing primarily relies on the \emph{sampling\_time} that encompasses the anneal-read cycle, iterated for a specified number of samples, as determined by the \textit{num\_reads} parameter (in our case this number is set to $1000$).

\subsubsection*{Solution Quality}

\begin{figure}[ht]
    \centering
    \includegraphics[width=1\linewidth]{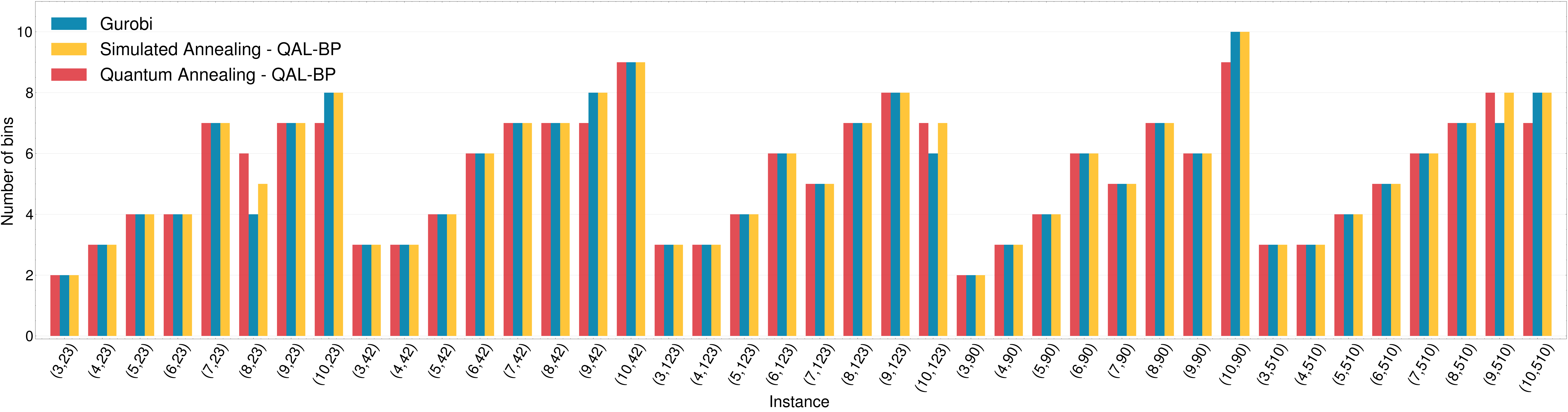} 
    \caption{Comparison of the number of bins corresponding to the solution with the minimum energy, for each problem instance, found by Gurobi, Simulated and Quantum annealing solvers.
    }
    \label{fig:num_bins}
\end{figure}

In terms of \textit{solution quality}, i.e., the number of bins used by a specific solution, Figure \ref{fig:num_bins} shows the results provided by the three solvers adopted. 


Gurobi solutions act as a benchmark, representing a well-established deterministic optimizer that correctly outputs the minimum required number of bins while explicitly taking into consideration all the constraints. The outcomes derived from SA validate the reliability of the QAL-BP formulation in consistently identifying feasible solutions for various problem instances. In the majority of cases, the optimal solution produced by SA aligns with the global optimum of the Gurobi. The only exceptions are noted in instances $(8, 23)$, $(10, 123)$, and $(9, 510)$, where the solutions generated by SA, while feasible, are characterized by a higher utilization of bins.

Concerning the outcomes derived from QA, it is evident that, in the majority of cases, the solutions align with those provided by SA. Deviations manifest in three potential scenarios. Firstly, the QA-generated solutions are suboptimal compared to SA, implying the utilization of a higher number of bins, as observed in $(8, 23)$. In such instances, the constraints imposed by the application of real quantum hardware lead to a feasible solution that, while not reaching the global optimum, remains within the bounds of feasibility. Secondly, instances of infeasible solutions may arise, exemplified by cases like $(10, 23)$, $(9, 42)$, $(10, 90)$, and $(10, 510)$, wherein the selected number of bins falls below the Gurobi solution, signaling the occurrence of overfilled bins. Indeed, because Gurobi explicitly considers the constraint on bin capacity, the number of bins yielded by it stands as the global optima for the given problems. Therefore, the lower count of bins in the QA solution indicates inadequacy, signifying an overfilled state for the utilized bins, as they fail to contain the items adequately. 
Thirdly, instances may arise where the selected number of bins appears to be appropriate based on the count of bins alone but proves to be infeasible due to violations of other constraints, e.g. instances $(10, 42)$ and $(8, 90)$ where, although the correct number of bins is selected, some of them have been overfilled. 
It is worth to notice that SA, employing the QAL-BP formulation, produces feasible solutions for the same problem instances. This substantiates that the infeasible solutions generated by QA are attributable to hardware errors rather than inherent deficiencies in the proposed formulation.



\subsubsection*{Feasibility Ratio}
\begin{figure}[ht]
    \center
    \includegraphics[width=0.8\linewidth]{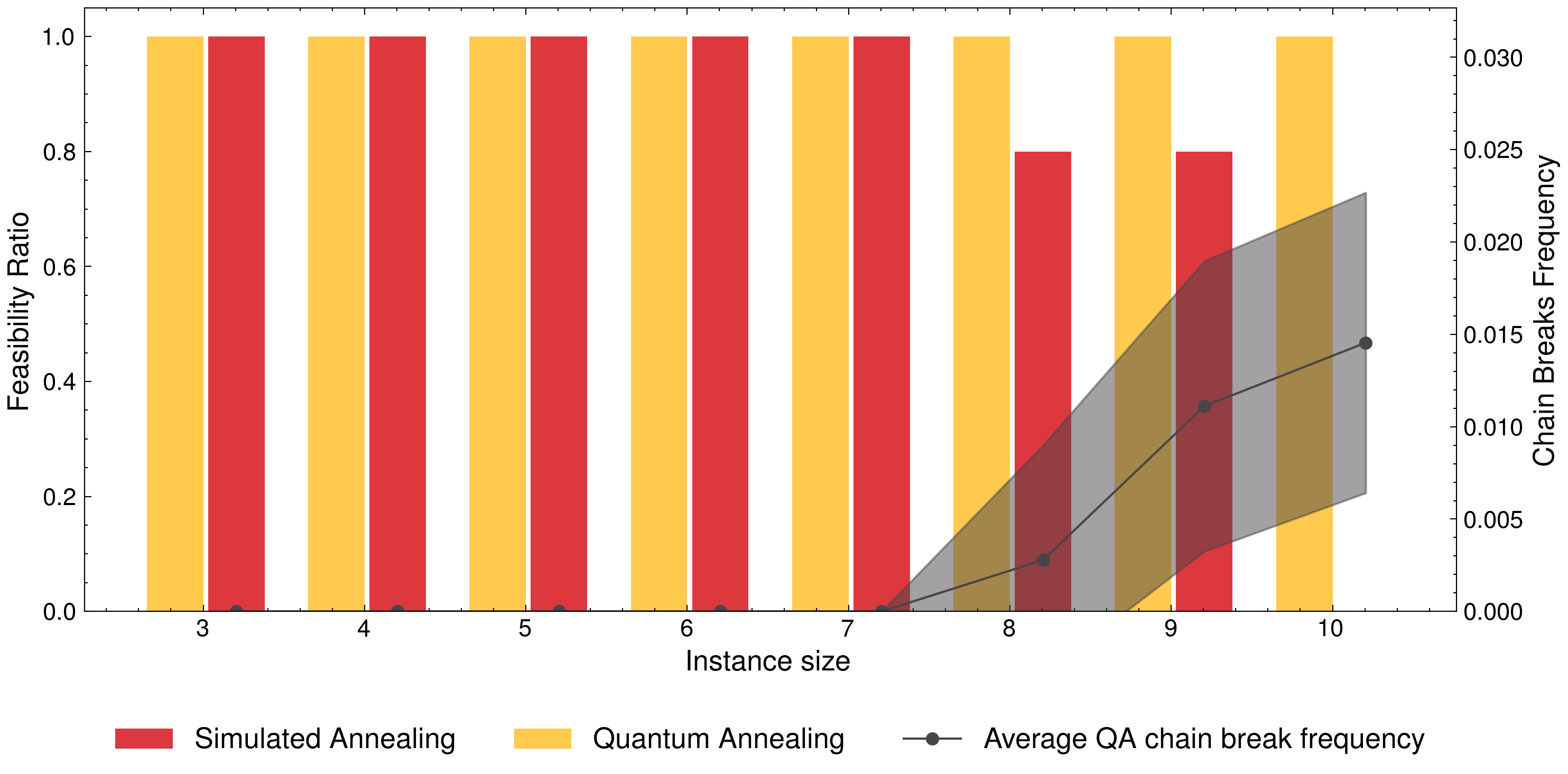}
    \caption{Comparison of the \emph{feasibility ratio} for each problem class. It is obtained as the ratio between the number of problems where the minimum found by QA is a feasible solution, divided by the number of problems with the same number of items.}
    \label{fig:feasible_density}
\end{figure}

To further highlight the challenges associated with the current QA hardware in accurately determining solutions for the QAL-BP formulation, Figure \ref{fig:feasible_density} depicts the \emph{feasibility ratio} across various instances of identical size. In the context of SA, solutions consistently manifest feasibility, as indicated by a \emph{feasibility ratio} of 1. This observation underscores that the QAL-BP formulation provides correct solutions for the BPP.

Concerning the QA implementation, for instances comprising up to $7$ items, QA uniformly yields feasible solutions in all cases. However, for instances involving $8$ and $9$ items, the \emph{feasibility ratio} diminishes to $80\%$. This reduction indicates that, in one out of the five instances tested with the same item count, the solution with minimum energy is infeasible.
In the 100\% of instances of 10 items, instead, 
QA was unable to find a feasible solution. Importantly, the observed decline in the \emph{feasibility ratio} appears to correspond to an increase in the \texttt{chain\_break\_fraction}.

In summary, the experimental findings illustrate that the QAL-BP formulation facilitates precise analytical estimation of penalty parameters, yielding near-global optimum solutions in all instances. However, the adoption of a real quantum annealer for the QUBO formulation of the BPP reveals suboptimal performances for medium-sized problems, attributable to existing limitations in current quantum hardware. Consequently, to fully harness the potential of the proposed QAL-BP method in effectively addressing larger instances of the BPP, imperative advancements in quantum technology and solver optimization are required.

\section*{Conclusion}

This paper introduces QAL-BP, a novel quantum formulation based on the Augmented Lagrangian method for efficiently solving the Bin Packing Problem (BPP) using quantum annealing. QAL-BP offers an analytical estimation of model's penalty terms for a specific class of problem instances, eliminating the need for recursive approximation methods to empirically estimate Lagrangian multipliers. This enhancement amplifies the generalizability of our approach to diverse input instances and improves efficiency by reducing the number of QUBO variables compared to alternative quantum formulations.

We demonstrated the effectiveness of our approach by solving larger problem instances than any previous QUBO formulation for the BPP. Additionally, we present the first experimental comparison of classical and quantum solutions for the BPP, validating that QAL-BP is an analytically correct QUBO formulation obviating the need for empirical estimation of penalty terms. Nevertheless, while our implementation on a quantum annealer does not outperform the state-of-the-art classical solver Gurobi, its TTS exhibits efficient scaling as the problem size increases, considering the current limitations of available quantum technology.

However, several limitations and challenges remain. Firstly, the generalizability of our model to generic BPP instances or other combinatorial optimization problems requires further investigation. Secondly, the limited number of qubits on current quantum annealers poses a significant challenge, restricting the size of problem instances that can be effectively solved. Consequently, testing our model on larger instances and evaluating scalability across a wider range of inputs is currently unattainable. Furthermore, noise and errors in the quantum annealer significantly impact the quality of provided solutions, particularly evident when dealing with larger problems, as demonstrated in experimental results compared to simulated annealing.

To address these challenges, future research will explore advanced quantum hardware with improved qubit accuracy and a greater qubits count. Another promising avenue involves investigating hybrid quantum annealing approaches that leverage classical and quantum methods in tandem, facilitating the solution of larger problem sizes beyond the capabilities of current QPUs. These endeavors are critical to further harnessing the potential of quantum computing in combinatorial optimization problems and propelling the field forward.

\bibliography{biblio}



\section*{Code Availability}

All code to generate the data, figures and analyses in this study is publicly available with detailed information on the implementation via the following repository: \href{https://github.com/Lorenz92/QAL-BP}{https://github.com/Lorenz92/QAL-BP}.

\section*{Acknowledgements (not compulsory)}

This work has been partially funded by the German Ministry for Education and Research (BMB+F) in the project
QAI2-QAICO under grant 13N15586.

\section*{Author contributions statement}


L.C. developed the theoretical formalism, performed the analytic calculations and numerical experiments. Both A.M. and M.L. supervised the project.
L.C. and A.M. wrote the manuscript in consultation with M.L.  
All authors provided critical feedback and helped shape the research, analysis, and manuscript.


\section*{Competing interests}

The authors declare no competing interests.






\end{document}